\documentclass{article}

\usepackage[numbers,square]{natbib}
\usepackage{amsmath,amssymb}
\usepackage{graphicx}
\usepackage{epstopdf}
\usepackage{amsthm}
\usepackage{picture}
\usepackage{color}
\usepackage{verbatim}
\usepackage{fancyhdr}
\usepackage{framed}
\usepackage[pdftex]{hyperref}
\usepackage{paralist}

\author{Wesam Elshamy}
\date{\today}
\title{Adaptive Control in Swarm Robotics}

\begin{document}

\maketitle

\begin{abstract}
Swarm robotic systems are mainly inspired by swarms of socials insects and the collective emergent behavior that arises from their cooperation at the lower lever.  Despite the limited sensory ability, computational power, and communication means of each swarm member, the swarm as a group manages to achieve difficult tasks such as searching for food in terrains with obstacles that individual robots cannot achieve in isolation of the other group members.  Moreover, such tasks are usually achieved without having information sharing capabilities at the swarm level or having a centralized decision making system.  In this report, I survey the state of the field of applying adaptive control method to increase swarm robotic systems robustness to the failure of individual robots, and increase its efficiency in performing its task.  A few techniques for the division of labor problem are briefly presented while one of them is  given in more detail.  A discussion of the advantages and disadvantages of this system is given and suggestions of potential improvements that can be made to the system are presented.
\end{abstract}

\section{Introduction}
Swarm robotic systems are mainly inspired by swarms of bees, flocks of birds or school of fish and the collective behavior that emerges in these groups due to the decentralized interactions between the group members \cite{dorigo99}.  The group can accomplish a task together, such as search for food, that the group members cannot perform independent from each other.

A swarm of robots is a group of robots that work together to perform a task.  Each member has a basic ability to sense its environment, like a temperature sensor or laser range viewer.  The input from such a sensor may not be very accurate.  A modest computational power and memory is available to each swarm member to process the sensory input, make the necessary calculations and reach a decision.  This computational power usually comes in the form of a micro chip built in the robot.  Each robot also has a limited ability to communicate with its peers.  The communication methods could be as simple as flashing a light or more sophisticated such as communicating using radio signals or using wireless computer networks.  The information the robot receives from its peers through its communication channels adds to its knowledge of the current state of the system and its environment.  One of the main distinctive features of swarm robotic systems over other systems of cooperating robots is the lack of global information sharing between robots; a robot can only communicate with robots in its vicinity using one or two-way communication channels.  Swarm robotic systems also lack a central decision making facility, or the ability of knowing the overall task progress.

A collective behavior similar to what has been observed in swarms of bees or flocks of birds has been attained in swarms of robots by following simple rules using sensory input and information acquired by communicating with other swarm members.   Such swarm robotics systems has been used to perform tasks such as searching hazardous environments \cite{mccarty09, kumar03}, foraging \cite{campo07}, exploration and mapping \cite{schmickl06, spears06}, self-organization \cite{shen04} and navigation \cite{nishimura11}.

Different adaptive control techniques are used in swarm robotic systems.  These techniques add robustness against the failure of individual robots, improve energy efficiency, help the swarm navigate unknown landscapes, get around obstacles and avoid getting trapped in dead-ends.  These systems has the benefit of being able to scale to a wide range of group sizes.  These techniques has to work in a distributed manner and has to rely on the limited capabilities each individual robot has in terms of environment sensing, computational power, and communication with other robots.

In this paper I survey the literature for different uses of adaptive control techniques in swarm robotic systems.  I also discuss the advantages and disadvantages of each one of them, and give a personal perspective on other potential applications for adaptive control methods in these systems.

\section{Current applications}
The main purpose of having an adaptive control mechanism in a system is to prevent its failure, maximize its efficiency, or simply make it work effectively in an environment that is dynamically changing.

In swarm robotic applications, failure may arise due to change in the robot's environment such as the terrain it is exploring or failure or malfunctioning of some of its sensors.  A swarm may fail to explore an unknown landscape if it gets stuck in a dead end or faced with an obstacle; an adaptive control system is needed to guide the swarm out the dead end and navigate around an obstacle.

Minimizing the energy used by a swarm of robots could be essential to the completion of an energy demanding task such as terrain exploration, search, foraging, or when the robots have limited capacity to store fuel or energy.  Optimizing energy use is also desirable for task completion cost reduction.

\subsection{Division of labor}
The collective foraging task is a well known and studied research problem in the field of swarm robotics due to its similarity to many real word problems.  In this task, a group of robots navigate their environment searching for food.  Food objects are randomly scattered in a restricted area in the robots' environment known as the foraging area.  Once a robot finds food, it carries it back to its home.  Real world applications of this task include toxic waste cleanup, harvesting and search and rescue \cite{cao97}.

In this collective foraging task, the group of robots performing the task consume energy while they are searching for food.  The energy consumed is related to the activities they perform.  For a simple foraging task where the robots have basic navigate tools and primitive communication means with its peers, the energy consumed is proportion to the distance the robot travels and to the amount of information it communicates with other robots.  The robots use the food they collect as their source of energy.  The net energy is the total energy acquired through food collection less any energy consumed by the robots.

The success of this task is measured by the group's energy efficiency.  Maximizing energy efficiency requires maximizing food collection while minimizing energy consumption.  Several factors come into play to achieve this end.  The number of robots actively searching for food has to be optimized.  Having too many active robots within a bounded foraging area leads to an unnecessary higher level of energy consumption as more robots are using up energy scouring their space for food.  Having more robots navigating a restricted area increases the chances of them meeting each other and communicating their status, which consumes more energy.  Moreover, interference among the robots decreases the robots chance of finding food.  On the other hand, if the number of actively foraging robots is very low, the group may not be able to collect enough food in a timely manner to make positive net energy, or even sustain its activities.  An optimal number of active robots is therefore needed to maximize the group's energy efficiency.  This number may even be dynamically changing with the environment as the amount of available food changes, and the foraging area boundaries move.  Due to that, the optimal number of active foragers may fluctuate over time to accommodate these changes and to optimize efficiency.

Researchers in the area of collective foraging followed different strategies.  Some of them focused on fining good communication mechanisms among robots to better understand the spacial characteristics of the foraging area.  This knowledge is then used in coordinating the robots efforts to search and find food \cite{werger96}.   Other researchers realized the effectiveness of trail-laying and following such as the one found in nature and used by ants and other insects \cite{sugawara02, shen02}

Other researchers focused their efforts on coming up with ways to divide the population of robots into two groups; one of them is actively searching for food, while the other group is waiting at home.  The objective of this division is to optimize the number of active foraging robots.  The task of dividing the robots into two such groups is known as \emph{division of labor}.  

\citet{labella04, cao97} introduced a simple adaptive method to change the ration of forager to waiting robots that improves the foraging performance.  \citet{jones03} Came up with a division of labor mechanism between collections of two different objects, while  \citet{guerrero03} invented an auction-like task allocation system to determine the optimal number of foragers needed for the foraging task.

\subsubsection{Example}
\citet{dai09} presented an adaptive division of labor model inspired by mechanisms used in social insects.  In his system, he developed a set of rules that assigns each robot $i$ a dynamic \emph{foraging probability} $P(i)$, where $i$ is the robot identifier number.  If this foraging probability value exceeds a threshold value $P_0$, the robot starts foraging, if not, the robot waits at home.

Two other variables are used in evaluating the foraging probability value; The foraging threshold $T_h(i)$ which relates to the foraging performance of robot $i$, and the foraging task stimulus $S$ which reflects the foraging stimulus by the group for this task.  Using these two variables, the foraging probability can be expressed as follows:

\begin{equation}
  \label{eq:1}
  P(i) = \frac{S^2}{S^2 + T_h^2(i)}
\end{equation}

Increasing the threshold value $T_h(i)$ decreases the chances that the robot will become a forager,  while increasing the incentive value $S$ makes it more likely to happen.

The values of $T_h(i)$ and $S$ in turn vary dynamically based on the robot's performance and interaction with its peers.  Each robot $i$ maintains a counter called \texttt{TaskCounter(i)} in which it records its observations of its peers.  If robot $i$ encounters a peer robot $j$ who has already found food, the robot increments its own \texttt{TaskCounter(i)} by one value.  If the peer robot is still searching for food, robot $i$ decrements its counter by a value of one.  Finally, if the peer has failed to find food, robot $i$ decrements its counter by a value of 2.

Every robot updates its foraging threshold value $T_h(i)$, and the global task stimulus value $S$ upon return to home.  The update values are based on the robot performance of achieving a positive (\emph{Success}) or negative (\emph{Failure}) net energy and based on its \texttt{TaskCounter(i)} value.  These update rules are shown in Table~\ref{tab:division_of_labor_rules}.

\begin{table}[th!]
  \centering
  \caption{Adaptation rules for division of labor algorithm}
  \begin{tabular}[l]{rl}
    \hline
    Success: &$\qquad T_h(i)\leftarrow T_h(i) - \Delta_1$\\
    Failuire: &$\qquad T_h(i)\leftarrow T_h(i) + \Delta_2$\\
    \hline
    Success and \texttt{TaskCounter(i)} $>$ 0: &$S \leftarrow S + \Phi_1$\\
    Failure and \texttt{TaskCounter(i)} $<$ 0: &$S \leftarrow S - \Phi_2$\\
    \hline
  \end{tabular}
  \label{tab:division_of_labor_rules}
\end{table}

Where $\Delta_1$, $\Delta_2$, $\Phi_1$ and $\Phi_2$ are model parameters that updates the robots threshold and task stimulus, respectively.

The first set of rules shown in Table~\ref{tab:division_of_labor_rules} increases the chances that a successful robot will take on another mission of foraging, and decreases such chances if the robot fails to make positive net energy.  The second set of rules updates the global incentives for the foraging task.  A successful robot with a positive \texttt{TaskCounter} means that the robot was able to find food and it encountered other robots who managed to collect food also.  This entails that there are plenty of food available for grasp and it increases the global foraging incentive $S$  by a value of $\Phi_1$.  On the other hand, a failed robot with a negative counter indicates a scarcity of food and that reduces the incentive by a value of $\Phi_2$.

\citet{dai09} tested this system with different parameter values and reported favorable results when compared to other systems that has fixed number and fixed ratio of active to waiting foragers.  The superiority of the system was more pronounced when the size of the swarm increased;  as this dynamic system adjusted the number of active foragers to reflect the availability of food and optimized its energy consumption, the other systems kept searching the environment without taking that factor into consideration leading to higher levels of energy consumption.

\section{Discussion}
The main objective of applying adaptive control techniques to swarm robotic systems is to increase their fault tolerance, improve their efficiency, and or to make them work efficiently.  The adaptive control methods presented in this report addresses one or more of these objectives with varying degrees.  Systems that emphasis fault tolerance tend to be less efficient as either redundant work is being done or idle workers are waiting for a malfunctioning swarm member to take its place.  In case of the division of labor example give earlier, idle members of the swarm were waiting at home for more food to be made available.  

Even though this method reduces the energy consumed by active workers, it ignores the energy needed to keep waiting members alive and the resources needed to maintain them.  A more efficient strategy would use swarm members that have different activity levels.  At higher levels, a swarm member will have better ability to collect food either by navigating the foraging area faster or by using longer laser range.  This higher activity or capability would consume more energy.  A swarm would be in this high level when its interaction with its peers indicate abundance of food.  On the other hand, when food is at normal levels, the robots would operate at normal levels of activity.  They would navigate their area at average speeds and use normal laser range to look for food.  This level of activity would consume energy less than the higher levels of activity.  Finally, a lower level of activity would be reached when food is scarce and it would require slower navigation speed and primitive food detection methods.  This lower level would consume the least amount of energy of all three levels.

\bibliography{bibliography}
\bibliographystyle{plainnat}

\end{document}